\documentclass[conference]{IEEEtran}
\IEEEoverridecommandlockouts

\usepackage{amsmath,amssymb,amsfonts}
\usepackage{algorithmic}
\usepackage{graphicx}
\usepackage{textcomp}
\usepackage{xcolor}
\def\BibTeX{{\rm B\kern-.05em{\sc i\kern-.025em b}\kern-.08em
    T\kern-.1667em\lower.7ex\hbox{E}\kern-.125emX}}
    
\usepackage[sorting=none]{biblatex}
\usepackage{gensymb}
\usepackage{soul}
\usepackage[markup=underlined]{changes}
\soulregister\cite7
\soulregister\ref7

\sethlcolor{lime}

\begin{document}

\title{Subtitle-based Viewport Prediction for 360-degree Virtual Tourism Video}

\newcommand{\orcidauthorB}{0000-0002-9592-0226}
\newcommand{\orcidauthorC}{0000-0003-2155-4507}

\author{\IEEEauthorblockN{1\textsuperscript{st} Chuanzhe Jing}
\IEEEauthorblockA{\textit{Graduate School of Engineering and Science} \\
\textit{Shibaura Institute of Technology}\\
Tokyo, Japan \\
mg20501@shibaura-it.ac.jp}
\and
\IEEEauthorblockN{2\textsuperscript{nd} Tho Nguyen Duc}
\IEEEauthorblockA{\textit{Graduate School of Engineering and Science} \\
\textit{Shibaura Institute of Technology}\\
Tokyo, Japan \\
nb20501@shibaura-it.ac.jp}
\and
\IEEEauthorblockN{3\textsuperscript{rd} Phan Xuan Tan}
\IEEEauthorblockA{\textit{Graduate School of Engineering and Science} \\
\textit{Shibaura Institute of Technology}\\
Tokyo, Japan \\
tanpx@shibaura-it.ac.jp}
\and
\IEEEauthorblockN{4\textsuperscript{th} Eiji Kamioka}
\IEEEauthorblockA{\textit{Graduate School of Engineering and Science} \\
\textit{Shibaura Institute of Technology}\\
Tokyo, Japan \\
kamioka@shibaura-it.ac.jp}
}

\maketitle

\begin{abstract}
360-degree streaming videos can provide a rich immersive experiences to the users. However, it requires an extremely high bandwidth network. One of the common solutions for saving bandwidth consumption is to stream only a portion of video covered by the user’s viewport. To do that, the user’s viewpoint prediction is indispensable. In existing viewport prediction methods, they mainly concentrate on the user's head movement trajectory and video saliency. None of them consider navigation information contained in the video, which can turn the attention of the user to specific regions in the video with high probability. Such information can be included in video subtitles, especially the one in 360-degree virtual tourism videos. This fact reveals the potential contribution of video subtitles to viewport prediction. Therefore, in this paper, a subtitle-based viewport prediction model for 360-degree virtual tourism videos is proposed. This model leverages the navigation information in the video subtitles in addition to head movement trajectory and video saliency, to improve the prediction accuracy. The experimental results demonstrate that the proposed model outperforms baseline methods which only use head movement trajectory and video saliency for viewport prediction.
\end{abstract}



\section{Introduction}

The high demand of 360-degree cameras in many fields such as robotics, virtual reality \cite{barmpoutis2020early,6dof}  
has led to the increasing of the global 360-degree camera market.
It has reached a value of 714 million USD in 2020 and is predicted to grow at an annual rate of around 25\% during 2021-2026  \cite{imarc}. The global VR headset market is expected to be above 44.2 billion USD by 2027 \cite{maximize}. These trends reflect a significantly growing demand for 360-degree video streaming which brings rich immersive experiences to the user by projecting the panoramic content on a virtual display. By wearing head-mounted displays (HMDs), such as Oculus Rift, HTC Vive, the user can obtain immersive experiences, feeling free to control his or her orientation during video playback. Currently, 360-degree videos can be available in 6K or even higher \cite{Qian2016}. However, delivering 360-degree videos with excellent Quality of Experience is challenging, due to restricted requirements of bandwidth and network latency \cite{Qian2016}.


Recent works have found that the bandwidth requirement can be reduced if only a small portion of the entire video frame viewed by the user, namely viewport, is delivered  \cite{He2018,Nasrabadi2019}.
This is  followed  by  tile-based  viewport  adaptive streaming - a promising approach in efficiently delivering 360-degree videos \cite{Xie2017,Ozcinar2019}. Instead of delivering the entire video frame with equally high quality, it only transmits high-quality tiles within the user’s viewport. Meanwhile, the other regions are transmitted in low quality or even discarded. 
Thus, viewport prediction plays a key role in such an approach. The state-of-the-art studies in viewport prediction can be categorized into trajectory-based \cite{Ban2018,Nasrabadi2020} and content-based \cite{Morais2021,Qiao2021} approaches. The former is based on the user’s history of head movement trajectory to predict the future viewport.
In this approach, the viewport prediction performance will be strongly influenced by the speed of the user's head movement, especially by the one in high motion videos. For example, in city guide videos where new scenes are introduced continuously, users tend to move their heads quickly and randomly to scan through all objects and stop at a certain region that they are interested in. This leads to the reduction of viewport prediction accuracy \cite{Yaqoob2020}.
The latter approach, on the other hand, relies on the video saliency to predict regions of interest that are more likely to attract the user's attention.
However, video saliency only depends on the visual content in a video. Meanwhile, there might exist additional video information, that can promisingly contribute to the improvement of viewport prediction.
For example, in city guide videos, navigation information that is explicitly or implicitly contained in either tour guide speech or video subtitles can suggest the user to locate and find the object (e.g., a church, a bridge) or special scene in a specific direction.
In this study, exploring the potential of navigation information in video subtitles to enhance viewport prediction accuracy is the main consideration.

In fact, beyond visual content in a video, subtitles are a considerably valuable addition \cite{Gernsbacher2015} providing user's viewport. Indeed, people prefer watching videos with subtitles even if they do not have to \cite{Gernsbacher2015}. The role of subtitles cannot be denied in a video for many reasons. For example, a quick search reveals that many people turn subtitles on when they are watching TV shows or movies, even if they are native speakers of the original language \cite{85-percent}. This is because they can understand video content better. Besides, there are over 28 million American adults who are deaf or hard of hearing \cite{Schiller2012}. If subtitles are not included in videos, a  huge audience will be excluded from the market. In addition, people might prefer to watch videos on mute\cite{85-percent}. Maybe they are in a public place and cannot be disruptive. Thus, it is hypothesized that when using a VR headset to watch 360-degree videos, the users will try to find or locate relevant objects or directions based on the information delivered in subtitles, which provides the possibility to enhance viewport prediction accuracy. Moreover, under the Covid-19 pandemic, virtual tourism videos have increasingly been becoming dominant choice on the Internet \cite{Bernd,Ramachandran}. In such a video type, subtitles usually contain a lot of navigation information which can be exploited for predicting the user's viewport.

Therefore, in this study, a subtitle-based viewport prediction model is proposed for 360-degree virtual tourism videos. The proposed model is a combination of three deep neural networks with dedicated responsibility, i.e., Convolutional Neural Network (CNN), Long Short-term Memory (LSTM), and Sequence-to-Sequence (Seq2Seq). The first two  networks, namely, CNN and LSTM, are utilized to extract video saliency and subtitle-based features, respectively. Then, these features along with the user's past head movement trajectory are combined and fed into the Seq2Seq network to generate the prediction of the user’s future viewport. The experimental results demonstrate that using subtitles, the proposed viewport prediction model outperforms baselines which only use head movement trajectory and video saliency as the input, in terms of prediction accuracy. The contributions of this paper are as follows:
\begin{itemize}
\item The hypothesis where the user's head movement trajectory is signiﬁcantly inﬂuenced by subtitles information is given and proved.

\item A deep subtitle-based viewport prediction model is proposed. This model can exploit navigation information in subtitles, in addition to saliency information and head movement trajectory, to provide more accurate viewport prediction. The proposed model is evaluated by comparing with baseline methods.
\end{itemize}

The rest of this paper is organized as follows: Section 2 provides related work and Section 3 describes the role of subtitles in a video. In Section 4, the proposed model for viewport prediction is introduced. The evaluation and experimental results of the model are presented in Sections 5. Finally, Section 6 concludes the paper.


\section{Related Work}

This section reviews some existing approaches in viewport prediction based on the input information that they used for the prediction. There are two commonly considered types of information: historical information and video content \cite{Yaqoob2020}. Historical head movement is often used by trajectory-based approaches to predict the future viewing position. These approaches leverage various methods, for example, 
KNN \cite{Ban2018}, clustering \cite{Nasrabadi2020}
and so forth.
Ban \cite{Ban2018} explored cross-users behaviors and user's personalized information concurrently by using K-Nearest-Neighbors(KNN) algorithm to predict long-term user's viewport. Further, Nasrabadi \cite{Nasrabadi2020} proposed a clustering-based viewport prediction approach that comprise past video streaming sessions as viewport pattern information. 
However, in these methods, long-term viewport prediction is highly inaccurate. The longer the prediction period, the less accurate the prediction is. At the same time, the prediction accuracy might be negatively influenced when the video contains high motion \cite{Yaqoob2020}. In such a video, the users usually move their heads quickly and randomly to look around before stopping at a certain region which they are interested in.

The latter information, in other words, video saliency, is used by content-based approaches to predict the user's viewport. 
Although plenty of research has been done for the detection of saliency maps on 2D images and videos, there are still many differences in the detection of VR or 360-degree images and videos. For example, to extract saliency features, 360-degree frames are usually converted to equirectangular (2D) images at first \cite{Zhang2018b}. 
Recently, some approaches \cite{Nguyen2018,Fan2017,Chen_2020} focus on extracting the saliency features in static 360-degree videos. Nguyen \cite{Nguyen2018} proposed a framework that combines saliency detection model while using head position tracking history for viewport prediction. 
Fan et al. \cite{Fan2017} developed a ﬁxation network to predict users’ head position by content-related saliency features. Video content information like motion is also used in the study of viewport prediction.
Chen\cite{Chen_2020} studied long-period prediction by concatenating video content-related features and the user's history head orientation information.
%
Since video saliency-based approaches only leverage the visual content in the video, the viewport prediction accuracy is still limited. Therefore, additional navigation information contained in the video should be explored to be considered for further accuracy improvement. In this study, video subtitles which implicitly and explicitly contain navigation information will be our main concern.


\section{THE ROLE OF SUBTITLE}

Subtitles mean to show the voice content of the video in the way of text. Because many words are homophonic, to better understand the video content, the users need the combination of subtitles and audio while watching videos. Users have grown accustomed to seeing text and video together. However, beyond that, subtitles also serve another arguably even more important purpose: accessibility \cite{Gernsbacher2015}. Subtitles make videos more accessible to a wider audience, including foreign-language speakers, hard-of-hearing individuals, and anyone who cannot watch a video with sound. When watching videos on public transportation or in a quiet library, people can completely turn off video sound and enjoy the video with subtitles. Adding subtitles to the videos is also an efﬁcient way to make the users, even native-speaking users, have a better understanding of video content. Figure 1 shows the display format and position of a subtitle in traditional 2D video and 360-degree video.

\begin{figure}[th]
\centerline{\includegraphics[width=\linewidth]{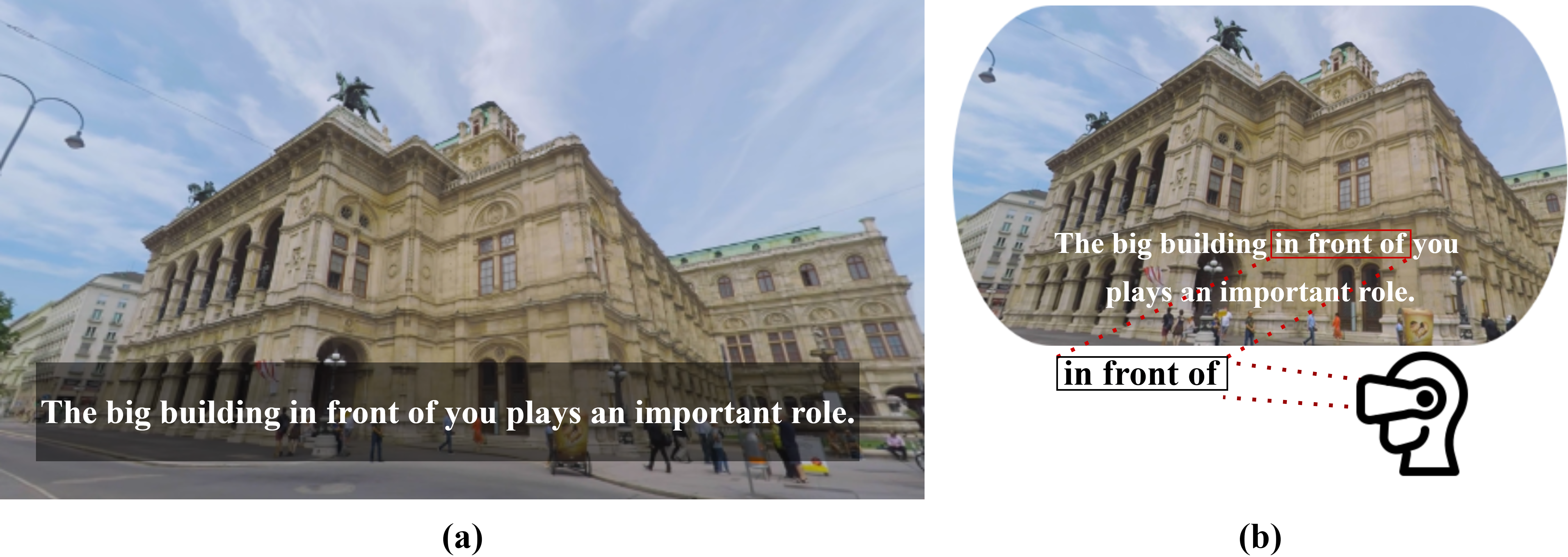}}
\caption{Subtitle in (a) traditional 2D video and (b) 360-degree video.}
\label{fig2}
\end{figure}

\subsection{Subtitles in traditional 2D video}

For traditional 2D videos, subtitles are usually located directly below the video as shown in Fig. 1a. People  have  the experience of reading subtitles, whether it is viewed on a personal electronic device or in a movie theater, the use of subtitles is essential. It means that people are very familiar with reading subtitles. Besides, people generally need to move their eyes to read the subtitles and watch the video with peripheral vision, which means people do not need to move their heads when watching. This is an important difference that can be found between subtitle in 2D videos and 360-degree videos.

\subsection{Subtitles in 360-degree video}

Some works \cite{Li2019,Brown2017} focused on the impact of subtitles in 360-degree videos and the behavior of users when watching 360-degree videos. Different from traditional 2D videos, the position and existence of subtitles in 360-degree videos are different in many ways due to the fact that the user’s viewport is changeable.

There are two forms of subtitle in 360-degree videos. The first one is a form where the subtitles are embedded in the video and will not move with the change of the user’s viewports \cite{360-subtitle}.  This form of subtitle is difﬁcult for the user to fully explore the video. The second one is a more common form where the subtitles are separated from the video and move accordingly with the change of the viewport, and will always be in a relative coordinate with the viewport. For example, the subtitles can always be in the lower center of the viewport as shown in Fig. 1b. The content (especially words for navigation information) of subtitles in 360-degree videos is also different from traditional videos. This is because the user’s viewport can only show part of the video and can be moved, so in 360-degree videos, the subtitle can tell the user how to move his or her head to ﬁnd the relevant objects by indicating the navigation words. For example, Fig. 1b also shows that when a user reads the subtitle of "in front of you", the user may turn around to find the building if it is not in the user's current viewport. Of course, subtitles sometimes do not use navigation words to guide the audience, but by directly introducing the relevant object or content to let the user move his or her head to explore.

\begin{figure*}[tb]

\centering
\includegraphics[width=0.95\linewidth]{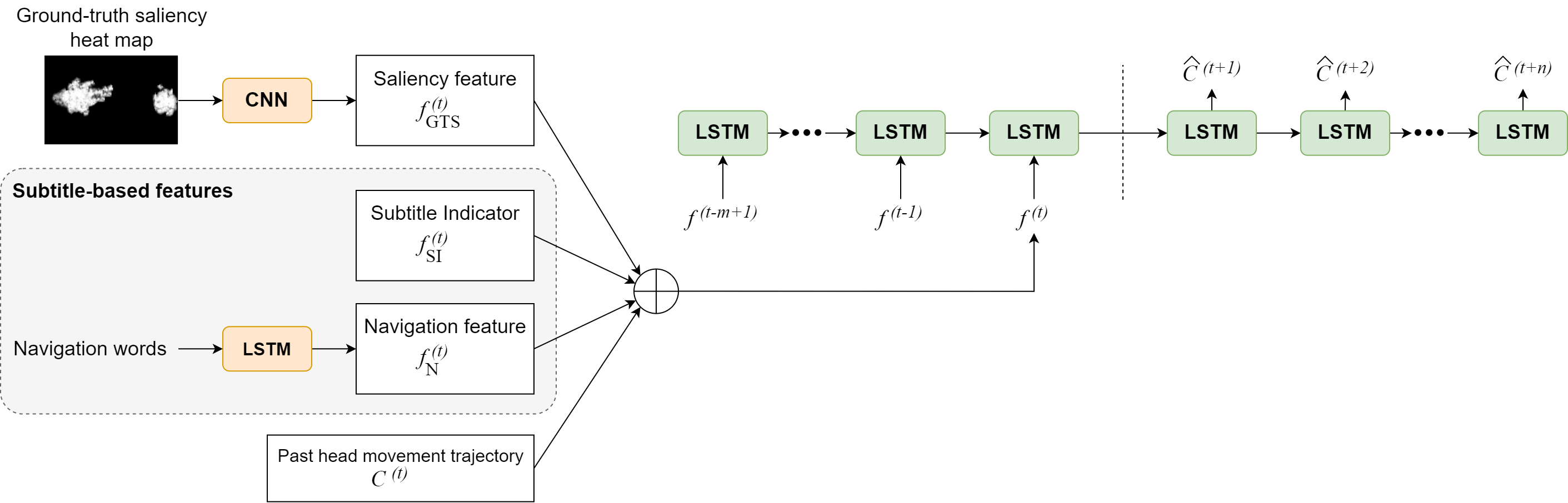}
\label{fig3}
\caption{Overall architecture of the proposed subtitle-based viewport prediction model.}
\end{figure*}  

\subsection{Subtitle in viewport prediction}

The 360-degree virtual tourism videos chosen in this study are mainly about city guidance. The second form of not embedded subtitle, which is separated from video and move follow with viewport, as described in Section 3.2, is used in the experiment. The subtitles can be downloaded separately from downloading videos. In this kind of tourism video, the tour guide will introduce the famous scenic spots such as statues, churches, or bridges in detail, so the information contained in the subtitles is of great value. When people read these meaningful sentences or words, they will instinctively want to ﬁnd the corresponding object. Therefore, the subtitle at a certain moment can be used to predict the viewport where people may watch in the next moment. 


\section{PROPOSAL}

In this section, the descriptions of input features and the architecture of the proposed viewport prediction model are presented. Fig. 2 shows the overall architecture of the proposed model. Accordingly, three kinds of features are utilized, i.e., saliency, subtitle-based, and past head movement trajectory features. These features form a high-dimensional feature vector which is used as the input of the Seq2Seq network to predict the user’s viewport. In the following subsections, the input features related to video saliency and video subtitles, and the architecture of the proposed model will be briefly described.

\subsection{Saliency feature}
In 360-degree videos, video saliency can be represented as a 2D heat map showing the most attractive regions in a video frame. Such regions have a high tendency to be seen by users, thus, they can be considered as the possible viewports \cite{Zhang2018b}. Therefore, video saliency plays an important role in the viewport prediction \cite{Nguyen2018}.
There are two approaches to extract video saliency, namely, content-based saliency and ground-truth saliency. The content-based saliency is predicted from the video content, which is highly dependent on the visual content \cite{MF2021,Zhang2018b}. On the other hand, the ground-truth saliency is extracted from collected users’ head trajectory, thus, it can be presented the viewing pattern of the users. It is important to note that in this study, we focus on 360-degree virtual tourism videos which contain a lot of navigation information. Such navigation information can deficiency of content-based saliency.
Therefore, the ground-truth saliency is considered to form saliency features for our proposed model.

To obtain the ground-truth saliency 2D heat map for a video frame at timestep \(t\) in video \(v\), the saliency value \(GTS_{v,x,y}^{(t)}\) at each heat map location \(P=(x, y)\) is calculated by the following equation \cite{MF2021}:

\begin{equation}
GTS_{v,x,y}^{(t)}=\frac{1}{N}\sum_{u}GTS_{v,u,x,y}^{(t)}, t\in [0,T]
\end{equation}

where \(N\) is the total number of users watching video \(v\), \(T\) is the time length of users’ head trajectory. \(GTS_{v,u,x,y}^{(t)}\) is a saliency value at heat map location \(P=(x, y)\) for user \(u\) in video \(v\) at timestep \(t\). It is obtained using a modification of radial basis function (RBF) kernel, which is also known as Gaussian function, as follows \cite{MF2021}:

\begin{equation}
GTS_{v,u,x,y}^{(t)}=exp\left ( -\frac{O(C_{v,u}^{(t)},P_{x,y}^{(t)})^2}{2\sigma ^2} \right )
\end{equation}
where, the \(C_{v,u}^{(t)}=(\varphi,\theta)\) is the center point of the user \(u\)'s viewport on video \(v\) at timestep \(t\). \(C\) is represented in the spherical coordinate system with azimuth angle \(\varphi\) and inclination angle \(\theta\). \(O(*)\) is the orthodromic distance between two points \(C_{v,u}^{(t)}\) and \(P_{x,y}^{(t)}\) in the surface of a unit sphere. \(\sigma\) presents the width of the Gaussian filter (which determines the degree of boundary smoothing). The larger \(\sigma\) is, the wider the band of the Gaussian filter and the better the degree of boundary smoothing. In this paper, \(\sigma=\pi/30\) is chosen to make the obtained ground-truth saliency map approximately similar with the saliency detection model \cite{Nguyen2018}.

The obtained ground-truth saliency heat map are input to a CNN network to extract a saliency feature vector \(f_{GTS}^{(t)}\) which will be used for viewport prediction in the proposed model.

\subsection{Subtitle-based features}
In order to leverage the subtitle information, two subtitle-based features are proposed, namely, (1) subtitle indicator and (2) navigation feature. Since not every video frame has a subtitle, the subtitle indicator \(f_{SI}^{(t)}\)is used as a binary indicator variable to indicate whether a subtitle is currently exists at timestep {$t$} or not. In other word,  \(f_{SI}^{(t)}=1\) when a subtitle exist at timestep \(t\), otherwise, its value is set to 0.

In tour-guide videos, a subtitle can contain a number of navigation words, e.g., “on the left”, “from the right”, “in front of”, etc. Such words can suggest the users to locate and look at the object or special scene in a specific direction. Therefore, we consider such kinds of word as a navigation feature, which promisingly provides those useful information to predict the user’s viewport. To extract such the information, an LSTM network with navigation words is utilized as an input. The output of the LSTM is the navigation feature vector \(f_N^{(t)}\) at timestep \(t\). If there is no subtitle at time step $t$ or the subtitle exists but it does not provide any navigation information, {$f_N^{(t)}$} will be equal to a vector of 0.

\begin{figure*}[tp]
\centering
\includegraphics[width=\linewidth]{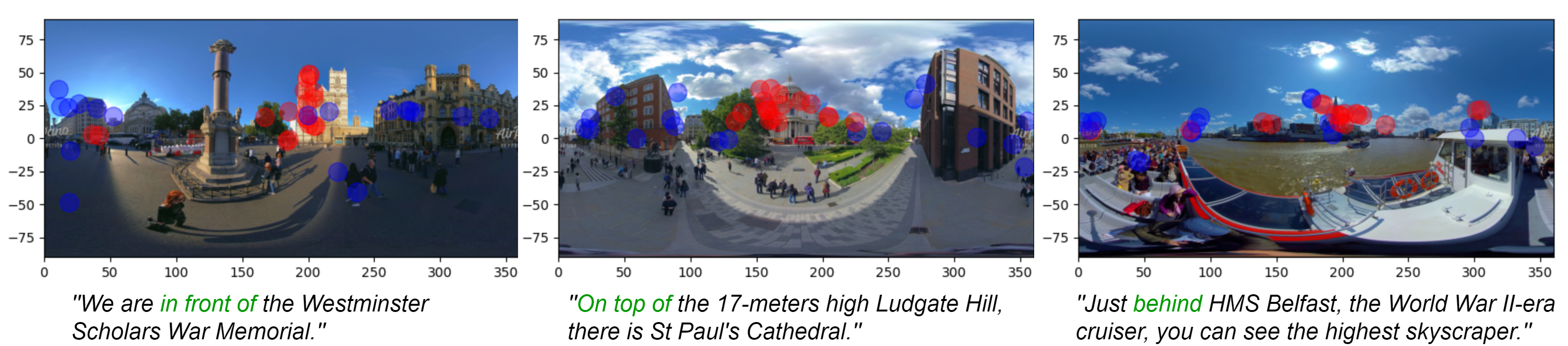}
\caption{Subtitle contains navigation words (e.g., "in front of", "on top of", etc.).}
\label{fig5}
\end{figure*}  


\begin{figure}[]
 \centerline{\includegraphics[width=\linewidth]{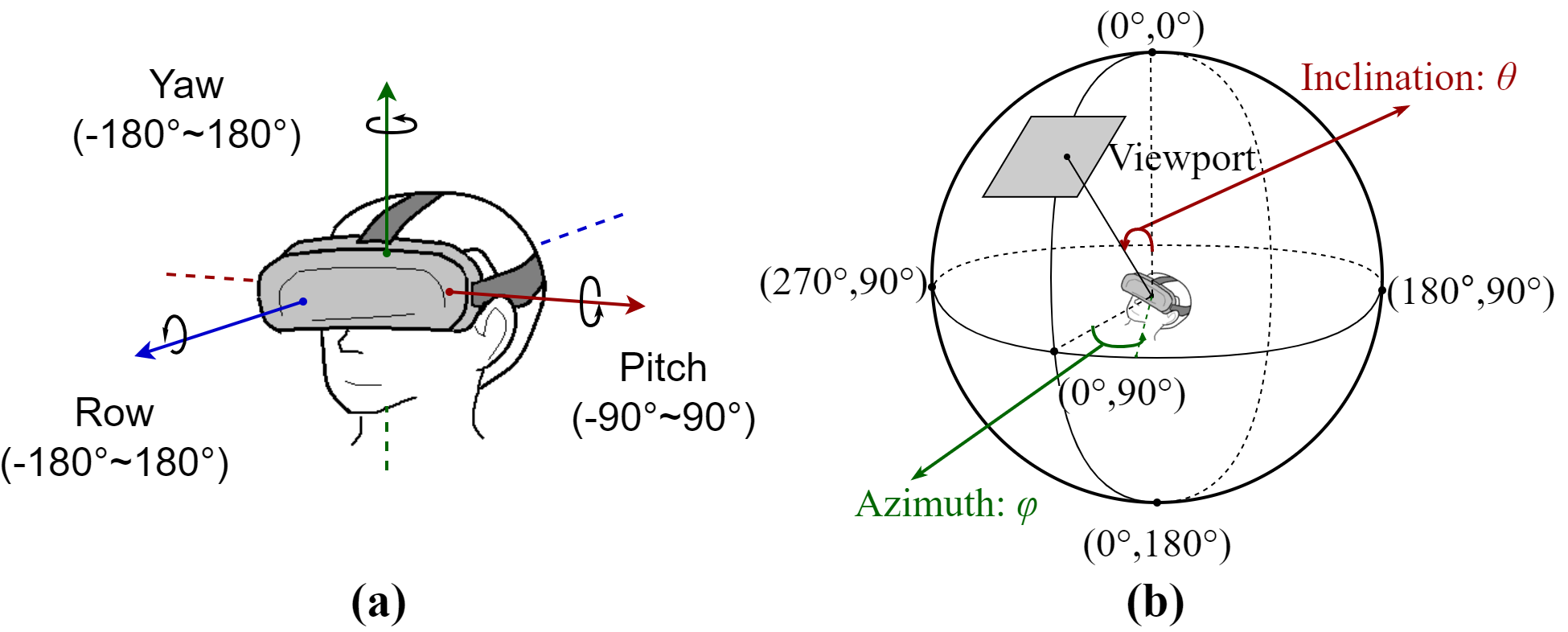}}
 \caption{User’s head trajectory in (a) Euler angles and (b) Spherical coordinate system.}
 \label{fig4}
\end{figure}

\subsection{Proposed viewport prediction model}
As mentioned in Subsection 4.1, the ground-truth saliency heat maps are fed into a Convolutional Neural Network (CNN) to extract saliency features \(f_{GTS}^{(t)}\). In the proposed model, the CNN network is based on VGG16 \cite{Simonyan2014}. The structure of VGG16 contains 16 layers and is mainly composed of 3x3 convolution kernel and 2x2 size of Max-pooling. The advantage of VGG16 is that stacking multiple small convolution kernels will increase the depth and width of the model, while slowing down the increase of computational amount. With 360-degree videos, the size of each video frame and ground-truth saliency heat map are very large. Therefore, to efficiently extract and process a saliency feature from a ground-truth saliency map, VGG16 is considered as an appropriate network for this study. In addition, navigation feature \(f_N^{(t)}\) is obtained by inputting navigation words to a multi-layered multi-unit LSTM network. The saliency feature \(f_{GTS}^{(t)}\), subtitle indicator \(f_{SI}^{(t)}\), navigation feature \(f_N^{(t)}\), and the user's past head movement trajectory $C^{(t)}$ are combined into a high-dimensional feature vector \(f^{(t)}=(f_{GTS}^{(t)},f_{SI}^{(t)},f_N^{(t)})\). This feature vector will be used as an input to the Seq2Seq network which is the main component in the proposed viewport prediction model. In the Seq2Seq network, the sliding window method [46] is used, where $m$ previous timestep feature vectors are taken as an input \((f^{(t-m+1)},…,f^{(t-1)},f^{(t)})\) and \(n\) ahead timesteps of user’s viewport \((\hat{C}^{(t+1)},...,\hat{C}^{(t+n)})\) are predicted. The reason of choosing Seq2Seq is that the model has strong ability to acquire sophisticated long-term dependencies among previous timesteps and provide a high performance prediction as shown in previous studies \cite{MF2021}.


\section{EVALUATION}

In this section, the hypothesis regarding the impact of subtitles on the user’s head movement is validated. Afterward, the performance and
computational complexity
of the proposed subtitle-based viewport prediction model
are
evaluated in comparison with baseline methods whose input features are video saliency and head movement trajectory.

\subsection{Data acquisition }

To validate the hypothesis and evaluate the {proposed} model’s performance, a subjective experiment was performed to explore users’ watching behaviors with subtitles and without subtitles. In this experiment, the participants were asked to watch a set of 360-degree virtual tourism videos using a Facebook Oculus Quest 2 VR headset. A video player in Unity \footnote{https://unity.com/}  was developed in order to collect the user’s head trajectory with the sampling interval of 0.5 seconds while they are watching the videos. 

The subjects’ head trajectory collected from Unity are presented in Euler angles as shown in Fig. 4a to describe their head rotation in 3D space. However, in order to get the viewport where the subject was looking at, those angles were converted to the spherical coordinate system as shown in Fig. 4b with the azimuth \(\varphi\) , inclination \(\theta\) and the subject's head position as the origin (0, 0).

\begin{table}[th]
\caption{List of videos used in the experiment}
\begin{tabular}{c c c c c}
\hline
\textbf{\textit{No.}} & \textbf{\textit{Name}}& \textbf{\textit{Length}}& \textbf{\textit{Subtitle duration}} &\textbf{\textit{Information}} \\
\hline
1 & London & 6min12s  & 202s& 4K,30fps \\

2 & Barcelona & 5min49s  &305s &4K,29fps \\

3 & Vienna & 4min32s  & 240s&4K,29fps \\

4 & Varadero & 3min29s  &178s &4K,29fps \\

5 & Holguin & 2min38s  &117s &4K,29fps \\
\hline
\end{tabular}
\label{tab1}
\end{table}

Five 360-degree virtual tourism videos: London \cite{Www.airpano.com2020}, Barcelona \cite{VRGorilla-VirtualReality&360Videos2020a}, Vienna \cite{VRGorilla-VirtualReality&360Videos2020b}, Cuba:Varadero \cite{VRGorilla-VirtualReality&360Videos2020c}, Cuba:Holguin \cite{VRGorilla-VirtualReality&360Videos2020d} as listed in Table 1, were used in the experiment. The total time length of subtitles is 17 minutes, and it is about 74\% of the total time length of ﬁve videos. 30 subjects were participated in the experiment. They were then equally divided into two groups, namely, Group A and Group B. Group A comprised of subjects who only watched the video with subtitles, whereas, subjects in group B only watched the video without subtitles. Each subject watched three different videos. In each group, each video was watched nine times. It is worth noting that before the experiment, a pre-survey was delivered to all the subjects. Accordingly, there were 16 subjects already experienced VR headsets before. Among them, there were 10 subjects used VR headsets to watch 360-degree videos. Therefore, in order to make sure that all the subjects are familiar with VR headsets and 360-degree videos, they were trained with several demo videos before starting the experiment. Demo videos were viewed directly on Youtube, and the duration ranged from three to ten minutes, depending on the subjects' first viewing experience. After training with demo videos, it is basically ensured that all subjects in each group have achieved similar VR experience. During the experiment, in order to eliminate the inﬂuence of video dubbing, the sound of all the videos was turned off.

\begin{figure}[h]
\centerline{\includegraphics[width=\linewidth]{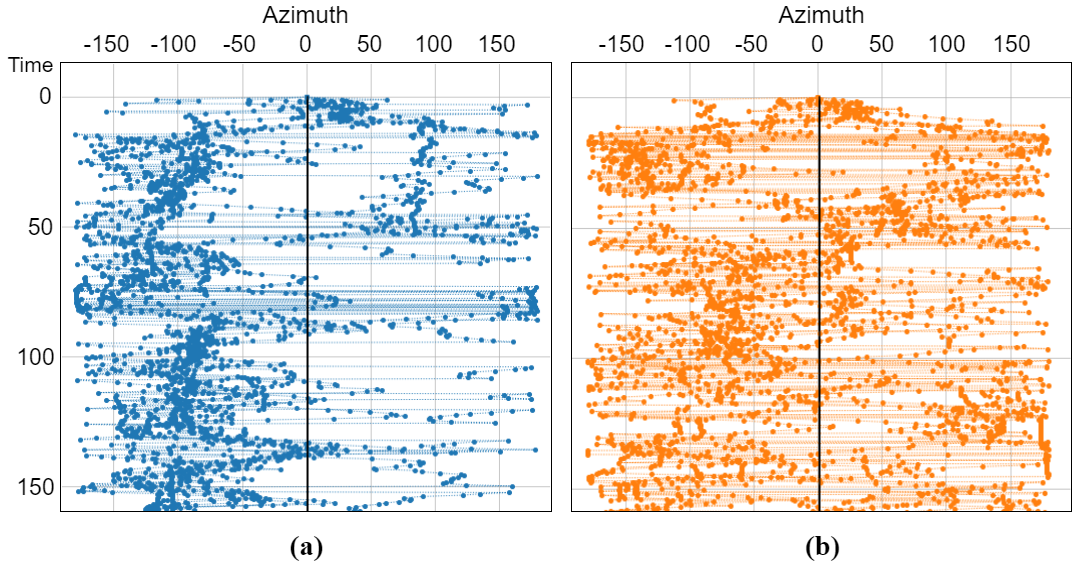}}
\caption{Head movement trajectory of two subjects from (a) Group A watching with subtitle) and (b) Group B (watching without subtitle).}
\label{fig7}
\end{figure}

\subsection{Impact of subtitle}

Fig. 3 show some examples of the viewport coordinates obtained from two groups of subjects. The red solid circles represent the viewport coordinates of the subjects in group A, who watched the videos with subtitles. Meanwhile, the blue solid circles represent the viewport positions of the subjects in group B, while they were watching videos without subtitles. Each illustrated video frame contains all the  18 subjects’ viewport coordinates data (9 from Group A and 9 from Group B) corresponding to one second.  Since the data was obtained twice a second, there are 18 solid circles of each color in every frame. In both figures, the red solid circles tend to cluster in the same regions in the frame. On the other hand, the blue solid circles seem to distribute randomly in different regions.

It means that the subjects in Group A who watched videos with subtitles, attempted to look for target regions in a particular frame. This is because the subtitles themselves include navigation information which comprises of directional words (e.g., “in front of”, “on top of”, etc) as shown in Fig. 3.
The viewport distribution of subjects in group B is more scattered due to the lack of effective navigation information. In other words,  their head movements are quite random.

To emphasize the role of subtitles, the difference between the distribution of head movements of two subjects from different groups is illustrated in Fig. 5. Obviously, with the support of navigation information in the subtitles, the subject in group A turned his or her attention to specific regions in the video frame. As the result, most of the blue dots in Fig. 5a gather in the same region. Meanwhile, a more random behavior is found in the head movements of the subject in group B, as shown in Fig. 5b. This result validates the hypothesis regarding the role of video subtitles in attracting the users’ attention during watching 360-degree videos. Based on this result, a further step is taken to validate the contribution of subtitles in viewport prediction, which will be presented in the next subsection.

\subsection{Performance of model}

In this section, the performance of the proposed model is evaluated by comparing with baseline methods. As mentioned in section 3, the proposed model considers the input features related to video saliency, subtitles, and head movement trajectory.

In order to provide a fair comparison, the proposed model was trained and tested on different set of videos that do not have the same content and user's viewing pattern. We used the data collected from 4 videos London \cite{Www.airpano.com2020}, Barcelona \cite{VRGorilla-VirtualReality&360Videos2020a}, Vienna \cite{VRGorilla-VirtualReality&360Videos2020b}, and Cuba:Varadero \cite{VRGorilla-VirtualReality&360Videos2020c} for training. The remaining video Cuba:Holguin \cite{VRGorilla-VirtualReality&360Videos2020d} was used for testing the performance of the proposed model. In the Seq2Seq network, we use $m=5$ previous timesteps as an input to predict $n=5$ ahead timesteps of user's viewport. The MSE was used as a loss function. The model was optimized using Adam with a learning rate of 0.001.

To validate the hypothesis of whether video subtitles have a noticeable contribution to viewport prediction and evaluate the performance of the proposed model, the baseline methods which only use video saliency and head movement trajectory for viewport prediction are considered. In this study, the following baseline methods were adopted for comparison with the proposed model:
\begin{itemize}
    \item MM18 \cite{Nguyen2018} used a Seq2Seq network to predict the user's future viewport. The network received two inputs:  (1) a ground-truth saliency heat map and (2) a mask which is encoded from the user's past head movement trajectory.
    
    \item NOSSDAV17 \cite{Fan2017} is employed to simply concatenate the ground-truth saliency heat map and past head movement trajectory as an input to a double-stacked LSTM. The output is then post-processed to produce the user's future viewport prediction.
    
    \item IEEENL20 \cite{Chen_2020} considered the user's head movement trajectory and content-related video frame for user's viewport prediction. The model leverages MobileNetV2 and double-stack LSTM to extract meaningful features from those two inputs.
\end{itemize}

\begin{table*}[th]
\caption{The viewport prediction accuracy of the proposed model in comparison with baseline methods.}
\begin{center}
\begin{tabular}{c c c c}
\hline
 & \textbf{\textit{$\text{RMSE}_\varphi$ (degree)}}& \textbf{\textit{$\text{RMSE}_\theta$ (degree)}} &\textbf{\textit{Average orthodromic distance}}\\
\hline
Proposed model & 58.6053 & 8.1483 & 0.3095 \(\pm\) 0.4601\\

\hline
MM18\cite{Nguyen2018} & 61.2725 & 8.6973 & 0.3929 \(\pm\) 0.3413\\
\hline
NOSSDAV17\cite{Fan2017} & 73.3164 & 15.7083 & 0.8389 \(\pm\) 0.35283\\
\hline
IEEENL20\cite{Chen_2020} & 63.5208 & 9.8963 & 0.6055 \(\pm\) 0.5940\\
\hline
\end{tabular}
\label{tab2}
\end{center}
\end{table*}

In addition, the following evaluation criteria were used in the evaluation process: 
\begin{itemize}
    \item Root mean square error (RMSE), which is capable of showing the spread out of the azimuth and inclination angles. Thereby, the errors between collected and predicted results of azimuth and inclination angles were calculated separately in this study. The values of RMSE of both azimuth and inclination angles were computed using the following equations: 

\begin{equation}
RMSE_\varphi = \sqrt{((\varphi-\hat{\varphi})^2)/2}
\end{equation}

\begin{equation}
RMSE_\theta = \sqrt{((\theta-\hat{\theta})^2)/2}
\end{equation}

where \(\varphi\) and \(\theta\) represents azimuth and inclination angles, respectively.

\item Orthodromic distance is the shortest distance between two points on the surface of a sphere by measuring along the surface of the sphere\cite{MF2021}. In this study, the Orthodromic distance was used to compute the distance between collected and predicted center points of users’ viewport. The Orthodromic distance between a predicted point \(\hat{C}(\hat{\alpha},\hat{\beta})\) and an collected point \(C(\alpha,\beta)\) was calculated as follows:

\begin{equation}
 \arccos({\sin(\hat{\alpha})\sin(\alpha)+\cos(\hat{\alpha})\cos(\alpha)\cos(\Delta\beta)})
\end{equation}

where, \(\alpha\) is the longitude and \(\beta\) is the latitude which are converted from spherical coordinate.
\end{itemize}

The lower azimuth, inclination RMSE, and orthodromic distance  values are, the more accurate the viewport prediction is.

Table 2 tabulates the performance of the proposed model in comparison with the baseline methods MM18 \cite{Nguyen2018}, NOSSDAV17 \cite{Fan2017} and IEEENL20 \cite{Chen_2020}. Accordingly, the proposed model provides relatively lower RMSE values compared to MM18, NOSSDAV17 and IEEENL20 in all cases, especially for inclination. At the same time, the lower orthodromic distance values can also be seen from the table.

\begin{figure}[tp]
\centerline{\includegraphics[width=\linewidth]{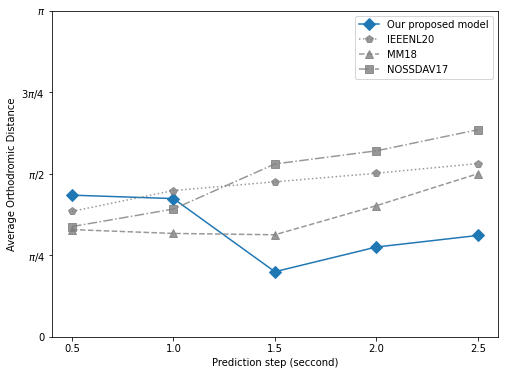}}
\caption{Average orthodromic distance of the proposed model and baseline methods under 10 prediction steps.}
\label{fig8}
\end{figure}

\begin{figure}[tp]
\centerline{\includegraphics[width=\linewidth]{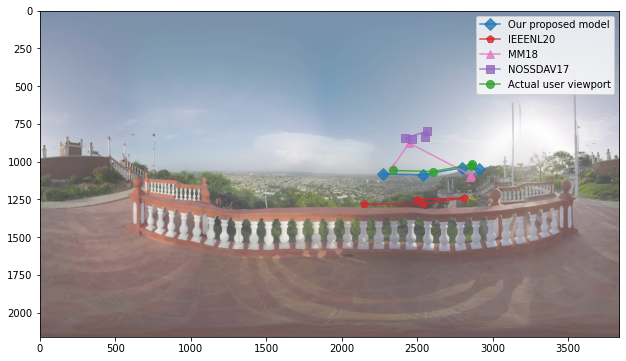}}
\caption{Visualization of predicted and actual viewports on a video frame.}
\label{fig9}
\end{figure}

Fig.6 illustrates the average orthodromic distance within 2.5-second prediction duration. In general, the proposed model produces lower average orthodromic distance compared to baseline methods. In some moments, the distance values obtained by the proposed model are noticeably higher than those of baseline methods, especially in the first two prediction steps. This is because, during that period, the video subtitles have not provided navigation information yet. Meanwhile, when the navigation information is shown from the prediction step at 1.5 second, the proposed model shows better results. Fig. 7 shows the qualitative evaluation where actual viewport and predicted viewports are altogether shown on a single sample video frame. It is obvious that the predicted viewport obtained from the proposed model tracks well with the actual one. This further emphasizes the contribution of video subtitles in the performance of viewport prediction. 

Once the video subtitles are introduced to the viewport prediction model, its direct and indirect contributions are recognized. The direct contribution can be seen when the features of video subtitle, namely, subtitle indicator and navigation features are considered as input features of the proposed model. The indirect contribution can be seen when the video subtitles influence the user’s head trajectory. Once  the coordinates of the user's viewing trajectory are collected in the experiments, the video subtitles will navigate the users to look for some specific content in a particular region in the video frame. Then these trajectory coordinates are used to generate ground-truth maps, which will be extracted as saliency features. In the previous studies \cite{Nguyen2018, Fan2017,Chen_2020}, their saliency maps are only generated from the video content, which does not necessarily coincide with the users' real viewport when the video subtitles exist. Therefore, this supports the idea which states that the saliency features are also influenced by the video subtitles. In the proposed method, it can be seen that the video subtitles indirectly influence the user’s head trajectory, which additionally contributes to the improvement of viewport prediction’s accuracy. 

\subsection{Computational complexity of model}

In order to verify the computational complexity of the proposed model, an experiment was conducted to measure the time that the model takes to predict $n$ ahead timesteps of user's viewport $(\hat{C}^{(t+1)},...,\hat{C}^{(t+n)})$. The timing experiment was carried out on a 20.04 Ubuntu LTS Intel i7-8750H @ 2.20GHz and 16GB RAM system. It is important to note that GPU devices were not used in this experiment. Despite that, the proposed model took only 597.4 milliseconds for prediction and 9.1 minutes to successfully train an epoch. Therefore, the proposed model is applicable for real-time user's viewport prediction in 360-degree video streaming.

\section{Conclusion}

In this study, the potential of video subtitles in the viewport prediction problem was carefully investigated. In fact, the navigation information in video subtitles efficiently supports the user in navigating the interesting objects in video frames. This results in a potentially predictable viewport during a 360-degree streaming session. Realizing the importance of video subtitles, in this study, we proposed a viewport prediction model in 360-degree video streaming, which takes into account the combination of video subtitles, video saliency, and head movement trajectory. To facilitate such a combination, our proposed model is comprised of CNN, LSTM, and Seq2Seq networks. The evaluation results showed that our proposed model achieved higher viewport prediction performance in comparison with baseline methods. This emphasizes the importance of combining the video subtitles with video saliency and head movement trajectory in viewport prediction. In other words, the navigation information provides a significant contribution to the improvement of prediction performance.




@article{Chen_2020,
	doi = {10.1109/lnet.2020.2977124},
  
	
  
	year = 2020,
	month = {jun},
  
	publisher = {Institute of Electrical and Electronics Engineers ({IEEE})},
  
	volume = {2},
  
	number = {2},
  
	pages = {81--84},
  
	author = {Xinwei Chen and Ali Taleb Zadeh Kasgari and Walid Saad},
  
	title = {Deep Learning for Content-Based Personalized Viewport Prediction of 360-Degree {VR} Videos},
  
	journal = {{IEEE} Networking Letters}
}

@article{barmpoutis2020early,
  title={Early fire detection based on aerial 360-degree sensors, deep convolution neural networks and exploitation of fire dynamic textures},
  author={Barmpoutis, Panagiotis and Stathaki, Tania and Dimitropoulos, Kosmas and Grammalidis, Nikos},
  journal={Remote Sensing},
  volume={12},
  number={19},
  pages={3177},
  year={2020},
  publisher={Multidisciplinary Digital Publishing Institute}
}

@InProceedings{6dof,
author="Attal, Benjamin
and Ling, Selena
and Gokaslan, Aaron
and Richardt, Christian
and Tompkin, James",
editor="Vedaldi, Andrea
and Bischof, Horst
and Brox, Thomas
and Frahm, Jan-Michael",
title="MatryODShka: Real-time 6DoF Video View Synthesis Using Multi-sphere Images",
booktitle="Computer Vision -- ECCV 2020",
year="2020",
publisher="Springer International Publishing",
address="Cham",
pages="441--459",
isbn="978-3-030-58452-8"
}

@misc{85-percent,
   author = {Sahil Patel},
   month = {May},
   title = {85 percent of Facebook video is watched without sound - Digiday},
   url = {https://digiday.com/media/silent-world-facebook-video/},
   year = {2016},
}

@article{360-subtitle,
  title={Subtitles in virtual reality: guidelines for the integration of subtitles in 360{\textordmasculine} content},
  author={Agull{\'o}, Bel{\'e}n and Matamala, Anna},
  journal={{\'I}kala, revista de lenguaje y cultura},
  volume={25},
  number={3},
  pages={643--661},
  year={2020},
  publisher={Escuela de Idiomas, Universidad de Antioquia}
}

@article{Schiller2012,
   author = {Jeannine S. Schiller and Jacqueline W. Lucas and Jennifer A. Peregoy},
   issue = {256},
   journal = {Vital and Health Statistics, Series 10: Data from the National Health Survey},
   pages = {1-80},
   pmid = {22834228},
   title = {Summary Health Statistics for U.S. adults: National Health Interview Survey, 2011},
   volume = {10},
   year = {2012},
}

@article{Gernsbacher2015,
   author = {Morton Ann Gernsbacher},
   doi = {10.1177/2372732215602130},
   issue = {1},
   journal = {Policy insights from the behavioral and brain sciences},
   month = {10},
   pages = {195},
   pmid = {28066803},
   publisher = {NIH Public Access},
   title = {Video Captions Benefit Everyone},
   volume = {2},

   year = {2015},
}

@article{Simonyan2014,
   author = {Karen Simonyan and Andrew Zisserman},
   journal = {3rd International Conference on Learning Representations, ICLR 2015 - Conference Track Proceedings},
   month = {9},
   publisher = {International Conference on Learning Representations, ICLR},
   title = {Very Deep Convolutional Networks for Large-Scale Image Recognition},

   year = {2014},
}

@article{MF2021,
author = {Romero Rondon MF and Sassatelli L and Aparicio-Pardo R and Precioso F},
doi = {10.1109/TPAMI.2021.3070520},
issn = {1939-3539},
journal = {IEEE transactions on pattern analysis and machine intelligence},
pmid = {33819149},
publisher = {IEEE Trans Pattern Anal Mach Intell},
title = {TRACK: A New Method from a Re-examination of Deep Architectures for Head Motion Prediction in 360-degree Videos},
volume = {PP},

year = {2021},
}

@misc{Ramachandran, 
author = {Dr. Selvakumar Ramachandran, Vijayalakshmi Subramani, Ivor Ambrose},
title = {{COVID-19 and opportunities for VR based tourism economy}},
url = {https://www.accessibletourism.org/?i=enat.en.news.2176},
urldate = {2021-08-20}
}

@misc{Bernd, 
author = {Bernd Debusmann Jr},
title = {{Coronavirus: Is virtual reality tourism about to take off? - BBC News}},
url = {https://www.bbc.com/news/business-54658147},
urldate = {2021-08-20}
}

@misc{maximize,
author = {Maximize Market Research Pvt.LTD},
title = {{Global Virtual Reality Headset Market – Industry Analysis (2020-2027)}},
url = {https://www.maximizemarketresearch.com/market-report/global-virtual-reality-headset-market/54754/},
urldate = {2021-08-20}
}

@misc{imarc,
author = {Imarc Group},
title = {{360-Degree Camera Market Size, Share, Industry Report & Forecast 2021-2026}},
url = {https://www.imarcgroup.com/360-degree-camera-market},
urldate = {2021-08-20}
}

@misc{VRGorilla-VirtualReality&360Videos2020d,
author = {{VR Gorilla - Virtual Reality & 360 Videos}},
month = {Oct},
title = {{Travel Cuba in 360 degrees VR - Episode 4: Holguin - 8K 360 VR Video - YouTube}},
url = {https://www.youtube.com/watch?v=sLwlVPHAbKs},
addendum = {accessed:2021-08-31},
year = {2020}
}

@misc{Www.airpano.com2020,
author = {www.airpano.com},
month = {Jul},
title = {{London, United Kingdom. Virtual travel. 360 video in 8K - YouTube}},
url = {https://www.youtube.com/watch?v=KGerjHMa90s},
urldate = {2021-08-31},
year = {2020}
}

@misc{VRGorilla-VirtualReality&360Videos2020a,
author = {{VR Gorilla - Virtual Reality & 360 Videos}},
month = {Jun},
title = {{Discover Barcelona In A Guided VR City Tour - 8K 360 VR Video - YouTube}},
url = {https://www.youtube.com/watch?v=EHEXgKjZ89M},
urldate = {2021-08-31},
year = {2020}
}

@misc{VRGorilla-VirtualReality&360Videos2020c,
author = {{VR Gorilla - Virtual Reality & 360 Videos}},
month = {Oct},
title = {{Travel Cuba in 360 degrees VR - Episode 5: Varadero and Trinidad - 8K 360 VR Video - YouTube}},
url = {https://www.youtube.com/watch?v=522-q9sSAFs},
urldate = {2021-08-31},
year = {2020}
}

@misc{VRGorilla-VirtualReality&360Videos2020b,
author = {{VR Gorilla - Virtual Reality & 360 Videos}},
month = {Feb},
title = {{A Guided City Tour of Vienna - 6K 360 VR Video - YouTube}},
url = {https://www.youtube.com/watch?v=BnVmijTniMU},
urldate = {2021-08-31},
year = {2020}
}

@article{Nasrabadi2020,
author = {Nasrabadi, Afshin Taghavi and Samiei, Aliehsan and Prakash, Ravi},
doi = {10.1145/3386290.3396934},
journal = {NOSSDAV 2020 - Proceedings of the 2020 Workshop on Network and Operating System Support for Digital Audio and Video, Part of MMSys 2020},
month = {Jun},
pages = {34--39},
publisher = {Association for Computing Machinery, Inc},
title = {{Viewport prediction for 360\° videos: A clustering approach}},

year = {2020}
}

@article{Li2019,
author = {Li, Ke and Yang, Di and Ji, Suhe and Liu, Liqun},
doi = {10.1109/ICIME.2018.00035},
journal = {Proceedings - International Joint Conference on Information, Media and Engineering, ICIME 2018},
month = {Jan},
pages = {130--134},
publisher = {Institute of Electrical and Electronics Engineers Inc.},
title = {{The Impacts of Subtitles on 360-Degree Video Journalism Watching}},
year = {2019}
}

@article{Brown2017,
author = {Brown, Andy and Schmitz, Anastasia and Turner, Jayson and Armstrong, Mike and Patterson, Jake and Glancy, Maxine},
doi = {10.1145/3084289.3089915},
journal = {TVX 2017 - Adjunct Publication of the 2017 ACM International Conference on Interactive Experiences for TV and Online Video},
month = {jun},
pages = {3--8},
publisher = {Association for Computing Machinery, Inc},
title = {{Subtitles in 360-degree Video}},

year = {2017}
}

@article{Zhang2018b,
author = {Zhang, Ziheng and Xu, Yanyu and Yu, Jingyi and Gao, Shenghua},
doi = {10.1007/978-3-030-01234-2_30},
journal = {Lecture Notes in Computer Science (including subseries Lecture Notes in Artificial Intelligence and Lecture Notes in Bioinformatics)},
month = {Sep},
pages = {504--520},
publisher = {Springer, Cham},
title = {{Saliency Detection in 360° Videos}},

volume = {11211 LNCS},
year = {2018}
}

@article{Qiao2021,
author = {Qiao, Minglang and Xu, Mai and Wang, Zulin and Borji, Ali},
doi = {10.1109/TMM.2020.2987682},
journal = {IEEE Transactions on Multimedia},
pages = {748--760},
publisher = {Institute of Electrical and Electronics Engineers Inc.},
title = {{Viewport-Dependent Saliency Prediction in 360° Video}},
volume = {23},
year = {2021}
}

@article{Morais2021,
author = {Morais, Dario D. R. and Althoff, Lucas S. and Prakash, Ravi and Carvalho, Marcelo M. and Farias, Myl{\`{e}}ne C.Q.},
doi = {10.2352/ISSN.2470-1173.2021.9.IQSP-255},
journal = {Electronic Imaging},
month = {May},
number = {9},
pages = {255--1--255--8},
publisher = {Society for Imaging Science & Technology},
title = {{A Content-Based Viewport Prediction Model}},
volume = {2021},
year = {2021}
}

@article{Ozcinar2019,
author = {Ozcinar, Cagri and Cabrera, Julian and Smolic, Aljosa},
doi = {10.1109/JETCAS.2019.2895096},
journal = {IEEE Journal on Emerging and Selected Topics in Circuits and Systems},
month = {Mar},
number = {1},
pages = {217--230},
publisher = {Institute of Electrical and Electronics Engineers Inc.},
title = {{Visual Attention-Aware Omnidirectional Video Streaming Using Optimal Tiles for Virtual Reality}},
volume = {9},
year = {2019}
}

@article{Xie2017,
address = {New York, NY, USA},
author = {Xie, Lan and Xu, Zhimin and Ban, Yixuan and Zhang, Xinggong and Guo, Zongming},
doi = {10.1145/3123266},
journal = {Proceedings of the 25th ACM international conference on Multimedia},
publisher = {ACM},
title = {{360ProbDASH: Improving QoE of 360 Video Streaming Using Tile-based HTTP Adaptive Streaming HTTP Server 360 video Source Cropping to tiles Segmentation and encoding}},

volume = {17},
year = {2017}
}

@article{He2018,
address = {New York, NY, USA},
author = {He, Dongbiao and Westphal, Cedric and Garcia-Luna-Aceves, J J},
doi = {10.1145/3229625},
journal = {Proceedings of the 2018 Morning Workshop on Virtual Reality and Augmented Reality Network},
publisher = {ACM},
title = {{Joint Rate and FoV adaptation in immersive video streaming}},
url = {https://doi.org/10.1145/3229625.3229630},
year = {2018}
}

@article{Nasrabadi2019,
archivePrefix = {arXiv},
arxivId = {1905.03823},
author = {Nasrabadi, Afshin Taghavi and Samiei, Aliehsan and Mahzari, Anahita and Farias, Mylene C. Q. and Carvalho, Marcelo M. and McMahan, Ryan P. and Prakash, Ravi},
doi = {10.1145/3304109.3325812},
eprint = {1905.03823},
journal = {Proceedings of the 10th ACM Multimedia Systems Conference, MMSys 2019},
month = {May},
pages = {273--278},
publisher = {Association for Computing Machinery, Inc},
title = {{A Taxonomy and Dataset for 360° Videos}},

year = {2019}
}

@article{Fan2017,
author = {Fan, Ching Ling and Lee, Jean and Lo, Wen Chih and Huang, Chun Ying and Chen, Kuan Ta and Hsu, Cheng Hsin},
doi = {10.1145/3083165.3083180},
journal = {Proceedings of the 27th ACM Workshop on Network and Operating Systems Support for Digital Audio and Video, NOSSDAV 2017},
month = {Jun},
pages = {67--72},
publisher = {Association for Computing Machinery, Inc},
title = {{Fixation prediction for 360° video streaming in head-mounted virtual reality}},
volume = {6},
year = {2017}
}

@article{Nguyen2018,
address = {New York, NY, USA},
author = {Nguyen, Anh and Yan, Zhisheng and Nahrstedt, Klara},
doi = {10.1145/3240508},
publisher = {ACM},
title = {{Your Attention is Unique: Detecting 360-Degree Video Saliency in Head-Mounted Display for Head Movement Prediction}},

year = {2018}
}

@article{Ban2018,
author = {Ban, Yixuan and Xie, Lan and Xu, Zhimin and Zhang, Xinggong and Guo, Zongming and Wang, Yue},
doi = {10.1109/ICME.2018.8486606},
journal = {Proceedings - IEEE International Conference on Multimedia and Expo},
month = {Oct},
publisher = {IEEE Computer Society},
title = {{CUB360: Exploiting Cross-Users Behaviors for Viewport Prediction in 360 Video Adaptive Streaming}},
volume = {2018-July},
year = {2018}
}

@article{Qian2016,
author = {Qian, Feng and Han, Bo and Ji, Lusheng and Gopalakrishnan, Vijay},
doi = {10.1145/2980055.2980056},
journal = {Proceedings of the Annual International Conference on Mobile Computing and Networking, MOBICOM},
month = {Oct},
pages = {1--6},
publisher = {Association for Computing Machinery},
title = {{Optimizing 360 video delivery over cellular networks}},

year = {2016}
}

@article{Yaqoob2020,
author = {Yaqoob, Abid and Bi, Ting and Muntean, Gabriel Miro},
doi = {10.1109/COMST.2020.3006999},
journal = {IEEE Communications Surveys and Tutorials},
month = {oct},
number = {4},
pages = {2801--2838},
publisher = {Institute of Electrical and Electronics Engineers Inc.},
title = {{A Survey on Adaptive 360° Video Streaming: Solutions, Challenges and Opportunities}},
volume = {22},
year = {2020}
}
\end{document}